# COMPARATIVE EVALUATION OF THREE METHODS OF AUTOMATIC SEGMENTATION OF BRAIN STRUCTURES USING 426 CASES

Mohammad-Parsa Hosseini[1*], Esmaeil Davoodi[3], Evangelia Bouzos[1], Kost Elisevich[2], Hamid Soltanian-Zadeh[3]

[1] Santa Clara University, Santa Clara, CA, USA
[2] Spectrum Health, Grand Rapids, MI, USA
[3] Henry Ford Health System, Detroit, MI, USA
mhosseini@scu.edu

**ABSTRACT**

Segmentation of brain structures in a large dataset of magnetic resonance images (MRI) necessitates automatic segmentation instead of manual tracing. Automatic segmentation methods provide a much-needed alternative to manual segmentation which is both labor intensive and time-consuming. Among brain structures, the hippocampus presents a challenging segmentation task due to its irregular shape, small size, and unclear edges. In this work, we use T1-weighted MRI of 426 subjects to validate the approach and compare three automatic segmentation methods: FreeSurfer, LocalInfo, and ABSS. Four evaluation measures are used to assess agreement between automatic and manual segmentation of the hippocampus. ABSS outperformed the others based on the Dice coefficient, precision, Hausdorff distance, ASSD, RMS, similarity, sensitivity, and volume agreement. Moreover, comparison of the segmentation results, acquired using 1.5T and 3T MRI systems, showed that ABSS is more sensitive than the others to the field inhomogeneity of 3T MRI.

## 1. INTRODUCTION

Automatic segmentation of brain structures in magnetic resonance imaging (MRI) has been a focus of attention in recent years due to the large datasets that need to be analyzed in different applications. Performing manual segmentation of these structures engages large datasets and is both costly and time consuming. This approach is also limited in its reproducibility but is considered the ground truth for its lesser susceptibility to imaging artifact and noise. Therefore, as automatic segmentation methods are developed, comprehensive comparison studies are needed to evaluate performance of these methods and investigate the advantages and disadvantages of each method as a suitable replacement for the manual approach. This will provide opportunity to refine these methods and adapt them for use in various conditions.

The hippocampus is considered as a key structure in studying temporal lobe epilepsy [1] and Alzheimer's disease [2]. In both conditions, the shape and volume of hippocampus has been used as a biomarker for diagnosis. To extract volumetric and shape features from MRI, proper segmentation of this structure is required. Because of its irregular shape, small size, and some poorly defined edges, automatic segmentation of this structure is challenging [3]. Hence such methods must be carefully scrutinized before being validated.

Nowadays, FreeSurfer [4] and HAMMER [5] are well-known tools for automatic segmentation. Prelabeled atlas images are mapped to the subject's images and then refined to label each part of the brain. Alternatively, other methods and software have evolved such as Automatic Brain Structure Segmentation (ABSS) [6] and LocalInfo [7] that are particularly designed to segment subcortical brain structures including the hippocampus. These methods use pattern recognition techniques in addition to registration algorithms to automate the segmentation process. They have outperformed FreeSurfer and HAMMER in a previous study [3, 31, 32] using MRI studies obtained from 195 cases. A limitation of that study which involved data acquired by both 1.5T and 3T MRI systems was not separately analyzed. Because field inhomogeneity artifact depends on the magnetic field strength [8] and is therefore different in 1.5T and 3T scanners the performance of individual automatic segmentation methods may have been variably affected.

In this work, we use MRI data obtained from 424 cases to evaluate comparative performance of three different automatic segmentation methods in segmenting the hippocampus. Four performance measures are calculated to evaluate the agreement of the automatic and manual segmentation results. We also investigate the effect of magnetic field strength (1.5T vs. 3T) on the performance of these methods.

## 2. LITERATURE REVIEW

Volumetric analysis of various brain structures is becoming increasingly common, as they can be used to diagnose diseases and monitor disease progression. The hippocampus is of importance, as it is implicated in several disease states, such as epilepsy and Alzheimer's disease. However, volumetric analysis of the hippocampus remains primarily manual. Automatic extractions are difficult due to poor definition of borders on MRI. Prior knowledge of the regional distribution, shape, atlas templates, among other aspects, are helpful in determining borders and to aid in the process of refining automatic segmentation methodology. Moreover, analysis of large cohorts is mandatory to overcome doubt pertaining to any unanticipated shortcomings of these methods.

Manual segmentation methods require a great deal of user interaction with commitment to all processing steps [9] to avoid transgressing borders inappropriately. With semiautomatic methods, users employ geometric shape and global localization constraints by marking points on the MRI or by drawing contours [9]. These methods can be based on Deformable Models or Atlases allowing for the use of geometric information already complied from real patient datasets to limit user interaction and reduce manual effort [9]. In fully automated methods, the statistical geometric information is obtained automatically from a reference atlas [9]. Some approaches use Single-Atlas, Multi-Atlas, or Probabilistic-Atlas. Others use deformable models or classifiers [9].

Deep convolutional neural networks (CNN) have emerged as an alternative to manual segmentation [33-35]. Ataloglou and colleagues [30] developed a CNN-based hippocampal segmentation (Hippodeep) that was compared to manual segmentation and FreeSurfer 6.0 [10]. This study was conducted with 200 healthy patients from seven locations across Canada [10]. The algorithm was shown to produce stable and reproducible data compared to other methods of segmentation. The approach requires additional testing in clinical populations to prove its promise for tracking volumetric changes in longitudinal studies [10].

In a joint international effort to harmonize existing manual segmentation protocols for the hippocampus, the European Alzheimer's Disease Consortium (EADC) and the Alzheimer's Disease Neuroimaging Initiative (ADNI) developed a Harmonized Protocol for Hippocampal Segmentation (HarP) [11]. The method developed is highly stable within and between human tracers and provides higher agreement than local manual segmentation protocols. However, only 20 MRI studies were segmented as this method was also very time consuming [11].

An ADNI dataset was used to evaluate a novel segmentation method based on iterative local linear mapping (ILLM). This method, which uses representative and local structure-persevered feature embedding, eliminates the need for registration [12]. It uses a semi-supervised deep autoencorder (SSDA) to exploit an unsupervised deep autoencorder and the local structure-preserved feature to transform the MR patch [12]. This method was tested with 135 subjects and was shown to be superior to the current patch-based multi-atlas segmentation (PBMAS) approach [12].

Semiautomatic methods, such as Fast Segmentation Through SURface Fairing (FASTSURF), are based on mesh processing techniques which use certain smoothness constraints to reconstruct hippocampal shape from a few manually delineated cross-sections of the hippocampus [13]. It was validated by comparing it to fully manual segmentation methods using different datasets and comparing it to FreeSurfer, an automated segmentation method [13]. FASTSURF results in highly accurate segmentations and require less effort in delineation than fully manual segmentation methods making it a promising tool [13]. Prior study of this method has used datasets of 12, 135 and 80 subjects [13] and deserves further confirmation with a larger cohort in other hands to promote its use.

Another semiautomated segmentation method (AdaBoost), was able to reliably analyze large datasets faster than a manual method [14]. MR images of 161 subjects taken from the ADCS Donepezil/Vitamin E clinical study were used to compare AdaBoost to manual segmentation [14]. Each method detected baseline differences in the hippocampus of patients who retained mild cognitive impairment throughout the clinical trial and those whose impairment progressed to Alzheimer's Disease [14].

An automatic segmentation method developed by Chupin, et al was created by adding hybrid constraints to an existing algorithm to make it fully automatic [15]. This method was based on the Iterated Conditional Modes (ICM) algorithm [15] and used global probabilistic atlas information to automatically initialize deformation and determine the bounding boxes and initial objects. It was better than semiautomatic segmentation and atlas-based segmentation in a small cohort of 16 patients [15]. FreeSurfer has been used as an automatic segmentation method in several papers. Research conducted by the University of Melbourne compared the utility of manual segmentation using two protocols for FreeSurfer and found that, although there was high reliability and agreement between protocols, more refinement was necessary and a dataset larger than one with 27 patients required [16]. FreeSurfer and FSL/FIRST were further compared to manual segmentation in quantifying hippocampal and amygdalar volumes with FreeSurfer found better correlated to manual segmentation [20]. Hippocampal segmentation by FreeSurfer-Subfields was also found superior to seven other automated methods in 105 ischemic stroke patients and 59 healthy participants [17]. However, when FreeSurfer was evaluated against manual tracing to determine hippocampal volume in a longitudinal study examining atrophy in middle and early old age, it was found to overestimate volume compared to manual segmentation [21]. Four segmentation methods: FreeSurfer, ANIMAL, nonlinear patch-based method, and nonlinear patch-based with and without error correction were compared using the ADNI dataset; the latter resulted in the most promising results [22]. In the pediatric population, FreeSurfer and FSL-First were compared to manual segmentation for use with the hippocampus and amygdala and neither were shown to have particularly strong agreement with manual segmentation [23].

The MAPS-HBSi method demonstrated superior reproducibility to other segmentation methods including manual segmentation, FreeSurfer, AdaBoost and FSL/FIRST when assessing atrophy rates of the hippocampus [18]. Another supervised algorithm, RUSBoost, showed a better dice index when compared to ADABoost, Random Forest, and FreeSurfer [19].

As witnessed above, there is substantial variance in the results comparing these various automated segmentation methods according to a variety of metrics applied and there is disagreement regarding the ultimate utility of any particular method, such as Freesurfer. Beyond this comes an assessment of subfield analysis in the course of disease or natural development as it may involve the hippocampus or other entity. A recent study analyzed the MRI of individuals aged 6 to 30 years [24]. Two semiautomated segmentation methods, Automated Segmentation of Hippocampal Subfields (ASHS) and Advanced Normalization Tools (ANTs), were compared to manual segmentation and showed that ASHS outperformed ANTs. Hippocampal subfield volume was also studied using ASHS in 32 patients addressing normal and early-stage nondementia Parkinson's disease [25]. Automatic segmentation methods of the hippocampus were also used to distinguish between Alzheimer's, mild cognitive impairment and normal aging [26].

. In previous work, the MRI studies of 195 patients were segmented by manual and automated methods, using LocalInfo, FreeSurfer, HAMMER, and ABSS [3]. ABSS was found to generate the most accurate results, followed by LocalInfo and FreeSurfer [3]. HAMMER scored the least accurate and was therefore excluded from this study.

In this work, we analyze a very large cohort using three automatic segmentation methods: ABSS, LocalInfo, and FreeSurfer. The methods are compared and evaluated on their performance in hippocampal segmentation using the MRI studies of 426 patients. Additionally, four performance measures are calculated to evaluate individual strengths against those of manual segmentation.

## 2. MATERIALS AND METHODS

### 2.1. Data

T1-weighted MRI study of 426 cases were used in this study. The data was acquired from 1993 to 2016 at Henry Ford Hospital, Detroit, MI from epilepsy patients (183 Males and 241 Females). The images were acquired using a 1.5T (154 of cases) or a 3T (270 of cases) MRI systems (GE Medical Systems, Milwaukee, USA / Philips MRI system, Best, The Netherlands) and T1-weighted imaging parameters were TR/TI/TE = 7.6/1.7/500 ms, flip angle = 20°, and voxel size = $0.781 \times 0.781 \times 2.0$ mm$^3$. On the 3T MRI, T1-weighted imaging parameters were TR/TI/TE = 10.4/4.5/300 ms, flip angle = 15°, and voxel size = $0.39 \times 0.39 \times 2.0$ mm$^3$.

### 2.2. Preprocessing

For all T1-weighted images, the brain was extracted from the scalp, skull, and dura using automated and manual skull stripping [3]. This procedure is not a prerequisite but improves the results of the automated segmentation methods used in this study. Both right and left hippocampi of each subject were manually segmented using MRIcro by a research assistant trained for this task [3, 27] and validated by a neuroscientist.

**2.3. Automated Segmentation**

In this study, we used three methods for automatically segmenting the hippocampi: ABSS, LocalInfo, and FreeSurfer.

**2.3.1 ABSS**

Automatic Brain Structures Segmentation (ABSS) is a pattern-recognition algorithm that is used for segmentation of brain structures in MR images. The ABSS method involves two stages. First, the shape of the desired brain structure is represented using geometric moment invariants (GMIs) in 8 scales. Each scale uses an artificial neural network (ANN) to approximate shape and signed distance functions of the structure. Input features of the ANN are the GMIs, voxel intensities and coordinates and the output feature is the signed distance function. Second, the outputs of all the ANNs in the first stage are combined and another ANN is designed to classify the voxels of the image based on whether they were inside or outside of the brain structure of interest [6]. In our study, we use this tool to train ANNs based on the shape (voxel intensities and coordinates) and signed distances of the hippocampal regions for the standard atlases or the user's training data. Finally, the subjects are processed and segmented based on the learned patterns of the training data. This package is freely accessible at [28].

**2.3.2 LocalInfo**

In the Local Information-Based Multiple-Atlas method (LocalInfo), multiple atlases are used to improve the likelihood of approximating an MR image with the desired atlas counterpart, resulting in improved segmentation. The desired atlas is selected for a specific brain region of interest and is utilized for location information extraction and for shape model. During initialization, nonrigid registration is used and a transformation is applied to the label maps. The most similar atlas label map is identified and both principal shapes and mean shapes are extracted. Additionally, a mean distance function for each structure is used for location information extraction. This is followed by a fine-tuning step, which is used to extract details of those brain structures that are not extracted from their primary shapes [7]. We use the LocalInfo method to segment the right and left hippocampi using information-based multiple atlases and affine registration to find the coordinates of each structure in each atlas mask. Finally, fuzzy classification, tissue-type information extraction, optimization of shape parameters, and transformation to label maps through nonrigid registration are employed for the segmentation task.

**2.3.3 FreeSurfer**

FreeSurfer [4] is an atlas-based segmentation tool designed for automated segmentation of brain structures including subcortical and cortical regions. It is a freely available [29] open source set of automated, robust, and accurate tools used to analyze neuroimaging data. It uses nonlinear alignment of the subject's image to the atlas labels (template matching) and then optimizes the transferred label map based on the subject's brain structures. To generate accurate models of various brain structures from MR images for segmentation, separate models for each structure at each point in space are generated. This enables the heterogeneity within a structure to be accounted for while keeping the distinction between pallidum and caudate, resulting in sharper and more informative distributions that make segmentation clearer. FreeSurfer also considers stereotypical spatial relationships between brain structures, by using the Markov Random Field (MRF) model and extending it to be spatially nonstationary. This allows for the ability to model the likelihood that the amygdala is above the hippocampus separately from the likelihood that it is below it. This creation of more sophisticated models enables the use of manually labeled training set images to bootstrap a whole-brain segmentation procedure.

**2.4. Segmentation Evaluation**

To evaluate the performance of segmentation methods, we calculated eight different performance measures by comparing the resulting label map of the hippocampi with the manually segmented hippocampi as the ground truth.
The *Dice coefficient* is defined as:

$$Dice\ coefficient = \frac{2N_{A \cap M}}{N_A + N_M} \tag{1}$$

where *A* and *M* are the voxels labeled as hippocampus by the automated and manual segmentation methods, respectively, and *N* represents the number of voxels in each set. This measure can have values in the range of 0.0 (i.e., no overlap) to 1.0 (i.e., complete overlap).

Precision is calculated as the ratio of the number of voxels *truly* labeled as the hippocampus, i.e., *true positive* (TP), to all voxels labeled as the hippocampus, which include the true positive voxels and the voxels labeled *falsely* as the hippocampus, i.e., *false positive* (FP) voxels,

$$Precision = \frac{TP}{TP + FP} \tag{2}$$

This measure also ranges from 0.0 (totally inaccurate detection) to 1.0 (fully accurate detection). It should be noted that a precision value of 1.0 does not necessitate a value of one for dice coefficient. As an example, assume half the hippocampus is labeled as hippocampus by a method. For this case, the precision will be 1.0 while the dice coefficient is equal to $1/1.5 = 0.667$. A related measurement is similarity, which includes FN, the voxel proportion of the hippocampus which segments as a background, in the denominator.

$$Similarity = \frac{TP}{TP + FP + FN} \tag{3}$$

Sensitivity is calculated using the following equation:

$$Sensitivity = \frac{TP}{TP + FN} \tag{4}$$

The *Hausdorff distance* is defined as:

$$H(A, M) = \max(h(A, M), h(M, A)) \tag{5}$$

where,

$$h(A, M) = \max_{a \in A} \min_{m \in M} \|a - m\| \tag{6}$$

where $a$ and $m$ are the coordinates of the points (voxels) on the automated (i.e., *A*), and manually (i.e., *M*), segmented structures, respectively. Here, $\|.\|$ denotes a norm function that can be the Euclidian distance. The Hausdorff distance decreases as the overlapping of the two structures increases.

The *rational absolute value degree* (RAVD) is based on the volume of the segmented structures,

$$RAVD = \frac{V_A - V_M}{V_M} \tag{7}$$

where *V* is the volume of the underlying structure. This measure is important in evaluating the methods when the biomarker is the volume of the structure. For example, in Alzheimer's Disease studies, where the volume of the hippocampus is one of the measures used to diagnose the disease [2], a segmentation method with lower RAVD value is more desirable.

The RMS measure is used to evaluate whether the quantity of segmentation varies or not by giving a distance metric using surface-to-surface geometries.

$$RMS(A, R) = \sqrt{\frac{1}{|S_A| + |S_R|} \times \left( \sum_{a \in S(A)} d^2(a, S_R) + \sum_{r \in S(R)} d^2(r, S_A) \right)} \tag{8}$$

$S_A$ is the set of automatic surface voxels, whereas $S_R$ is the set of manual surface voxels. The equation, $d^2(a, S_A)$, is the nearest Euclidian distance from a point on the surface (a) to the surface ($S_A$) squared.

A related measure is ASSD, which is the average distance of all the contour points that have been automatically segmented to the closest contour point that has been manually traced. The notations for ASSD are the same as the ones for RMS.

$$ASSD(A, R)^n = \frac{1}{|S_A| + |S_R|} \times \left( \sum_{a \in S(A)} d(a, S_R) + \sum_{r \in S(R)} d(r, S_A) \right) \tag{9}$$

## 3. RESULTS AND DISCUSSION

FreeSurfer, ABSS, and LocalInfo were used to automatically segment the hippocampal structure of 426 T1-weighted MRI studies. The eight performance measures discussed in Section 2 were calculated to compare the automatic segmentation results with the manually segmented hippocampi. Table 1 shows performance measures from 6 patients as a representation of the entire dataset. In Fig. 1 and 2, the boxplots of these measures are shown and in Fig. 3 and 4, the scatterplots of the data are shown. Table 2 shows that ABSS results for hippocampal volumes approximate more closely the results from manual segmentation. Table 3 is a standard ANOVA table that shows a p value of < 0.05; the null hypothesis is rejected identifying the three methods of automatic segmentation as having significantly different results.

| Subject | Method | Hausdorff | Dice | Similarity | Precision | RMS | ASSD | MeanDistance | Sensitivity | RAVD |
|---|---|---|---|---|---|---|---|---|---|---|
| 1 | ABSS | 1.00 | 0.86 | 0.76 | 0.84 | 0.56 | 0.29 | 0.32 | 0.89 | 0.07 |
|   | LocalInfo | 3.00 | 0.72 | 0.56 | 0.63 | 1.23 | 0.74 | 0.92 | 0.84 | 0.34 |
|   | Freesurfer | 3.46 | 0.70 | 0.54 | 0.57 | 1.35 | 0.84 | 1.08 | 0.93 | 0.63 |
| 2 | ABSS | 1.41 | 0.83 | 0.72 | 0.89 | 0.69 | 0.38 | 0.31 | 0.78 | -0.12 |
|   | LocalInfo | 2.83 | 0.73 | 0.58 | 0.68 | 1.11 | 0.70 | 0.82 | 0.79 | 0.17 |
|   | Freesurfer | 3.74 | 0.70 | 0.54 | 0.58 | 1.39 | 0.86 | 1.11 | 0.90 | 0.54 |
| 3 | ABSS | 4.00 | 0.74 | 0.59 | 0.71 | 1.36 | 0.72 | 0.90 | 0.78 | 0.10 |
|   | LocalInfo | 5.92 | 0.61 | 0.44 | 0.52 | 2.18 | 1.31 | 1.73 | 0.75 | 0.44 |
|   | Freesurfer | 7.07 | 0.64 | 0.48 | 0.56 | 2.49 | 1.40 | 1.73 | 0.77 | 0.38 |
| . | ABSS | . | . | . | . | . | . | . | . | . |
| . | LocalInfo | . | . | . | . | . | . | . | . | . |
| . | Freesurfer | . | . | . | . | . | . | . | . | . |
| 424 | ABSS | 1.41 | 0.85 | 0.73 | 0.83 | 0.86 | 0.43 | 0.47 | 0.87 | 0.05 |
|   | LocalInfo | 6.40 | 0.74 | 0.59 | 0.64 | 2.20 | 1.04 | 1.44 | 0.87 | 0.35 |
|   | Freesurfer | 5.10 | 0.76 | 0.62 | 0.67 | 1.77 | 0.89 | 1.15 | 0.88 | 0.31 |
| 425 | ABSS | 1.00 | 0.86 | 0.75 | 0.86 | 0.59 | 0.30 | 0.31 | 0.85 | -0.01 |
|   | LocalInfo | 3.61 | 0.72 | 0.56 | 0.64 | 1.33 | 0.78 | 1.01 | 0.82 | 0.29 |
|   | Freesurfer | 4.12 | 0.65 | 0.48 | 0.52 | 1.57 | 1.02 | 1.33 | 0.87 | 0.68 |
| 426 | ABSS | 4.24 | 0.73 | 0.58 | 0.72 | 1.48 | 0.87 | 1.02 | 0.74 | 0.03 |
|   | LocalInfo | 5.39 | 0.68 | 0.52 | 0.59 | 1.92 | 1.19 | 1.55 | 0.80 | 0.35 |
|   | Freesurfer | 5.00 | 0.69 | 0.53 | 0.58 | 1.81 | 1.10 | 1.46 | 0.86 | 0.47 |

**Table 1:** Performance comparisons of three automatic segmentation methods using eight metrics conducted on 426 patient samples.

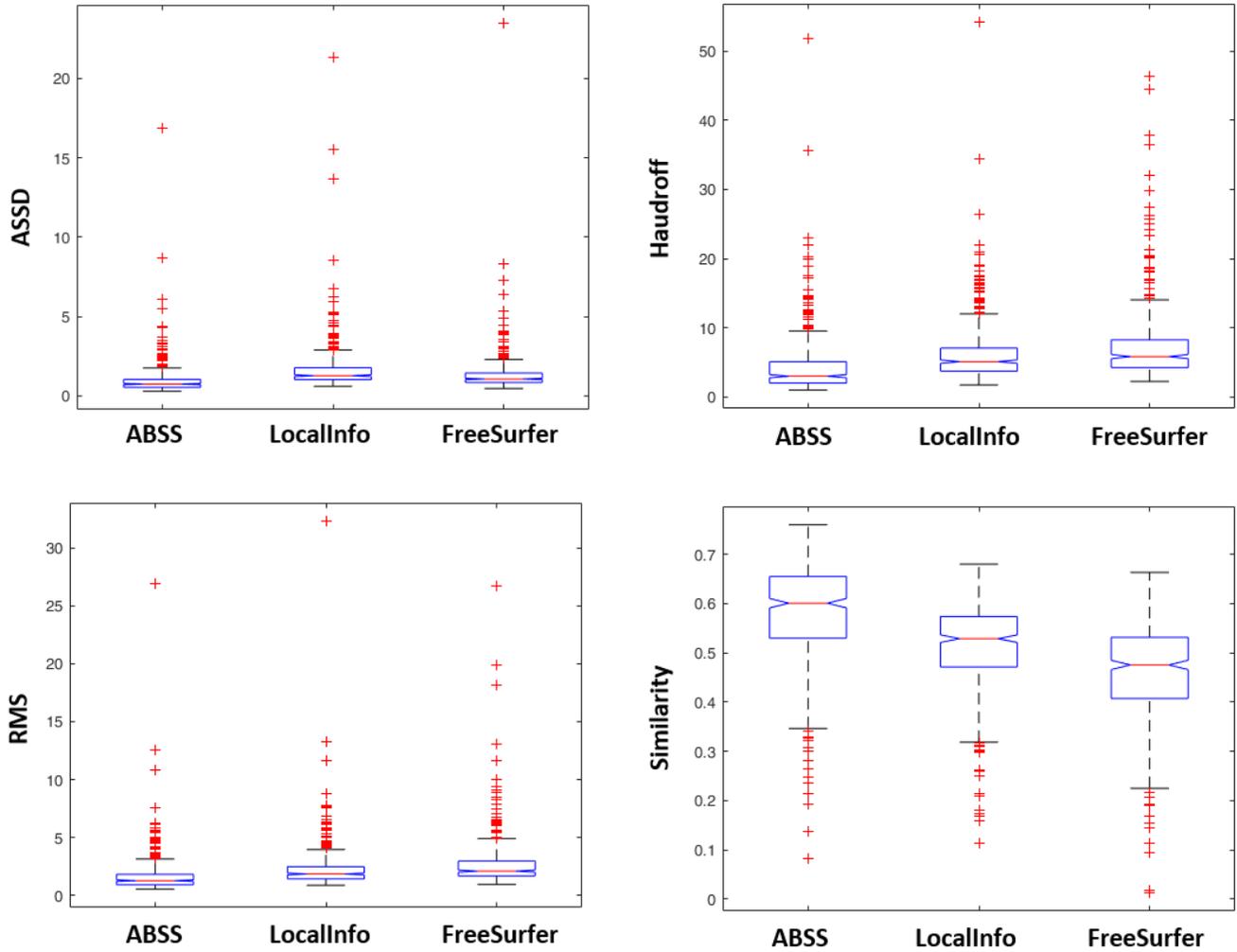

**Figure 1:** Comparison of the performance of the three automated methods to segment the hippocampus using ASSD, Haudroff, RMS, and Similarity as metrics. The boxplots of the four segmentation evaluation measures are shown for each method. Overall ABSS outperforms LocalInfo and FreeSurfer.

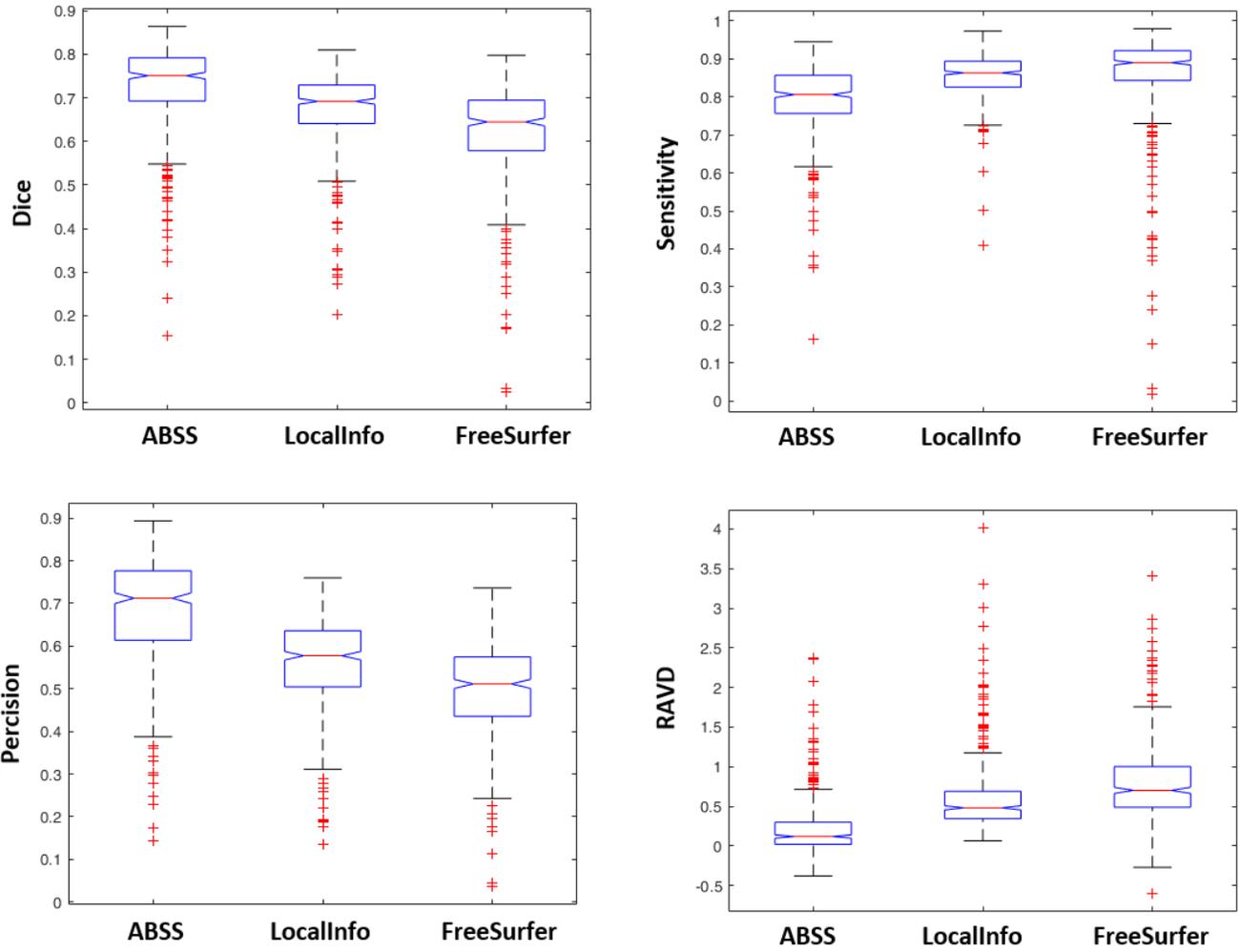

**Figure 2:** Comparison of the performance of the three automated methods to segment the hippocampus with Dice, Sensitivity, Precision and RAVD as metrics. The boxplots of the four segmentation evaluation measures are shown for each method. Overall, ABSS outperforms LocalInfo and FreeSurfer.

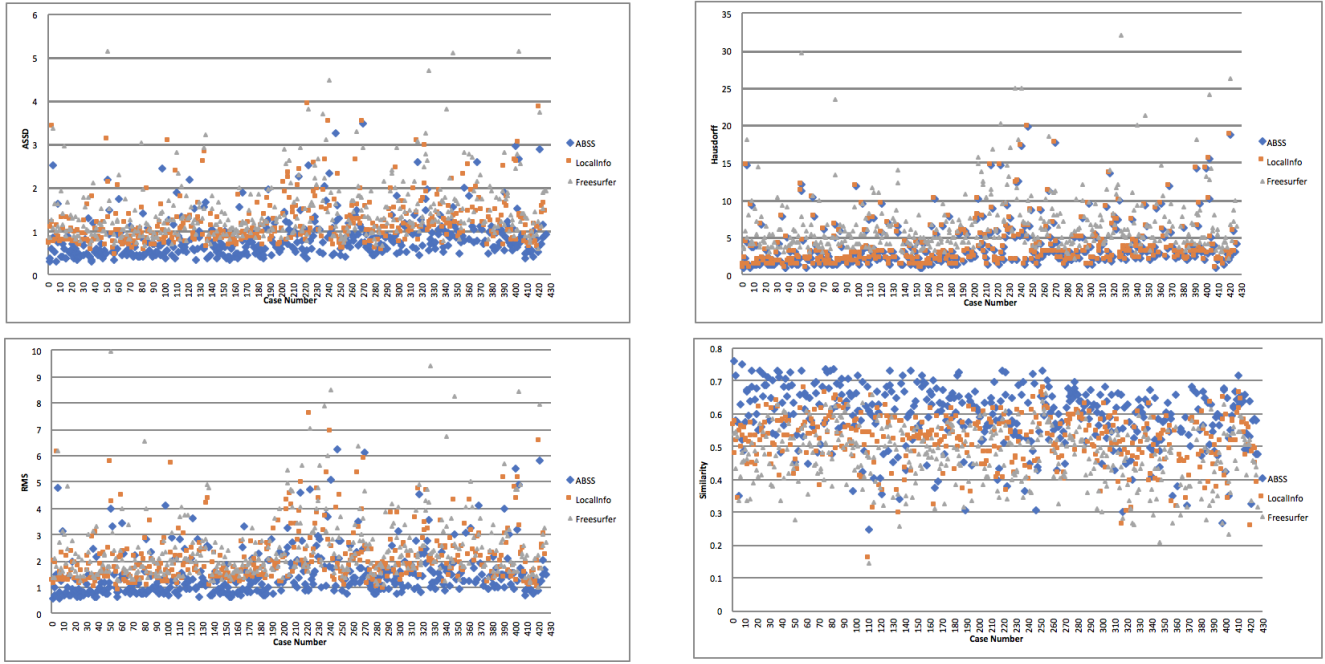

**Figure 3:** Scatterplots of ASSD, Hausdorff, Similarity, and RMS using the three segmentation methods of 426 patient samples.

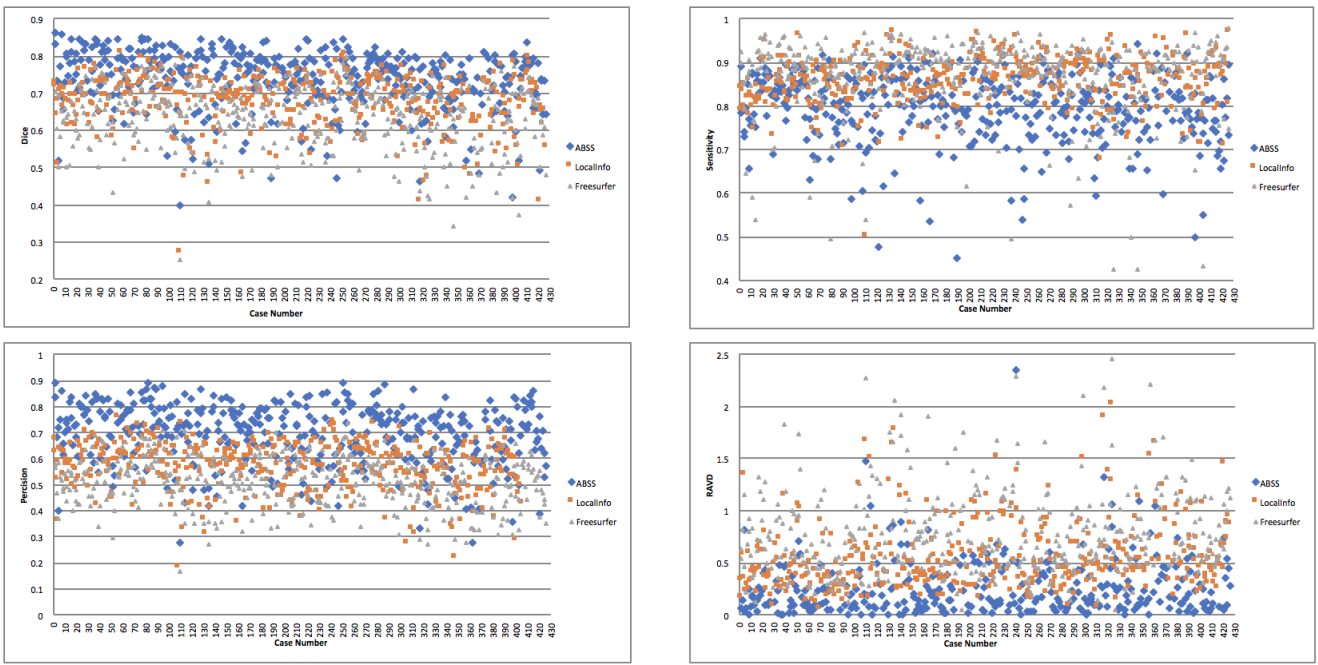

**Figure 4:** Scatterplots of Dice, Sensitivity, Precision, and RAVD using the three segmentation methods in 426 patient samples.

| Subject | Method | Left Volume (Voxel) | | | Right Volume (Voxel) | | |
|---|---|---|---|---|---|---|---|
| | | Automatic Segmentation | Manual Segmentation | Normalization of Difference in Volume | Automatic Segmentation | Manual Segmentation | Normalization of Difference in Volume |
| 1 | ABSS | 3318.19 | 2863.16 | 0.16 | 3010.77 | 3076.64 | 0.02 |
| | LocalInfo | 4114.80 | 2863.16 | 0.44 | 3824.46 | 3076.64 | 0.24 |
| | Freesurfer | 4849.19 | 2863.16 | 0.69 | 4841.87 | 3076.64 | 0.57 |
| 2 | ABSS | 2921.71 | 3114.46 | 0.06 | 2643.57 | 3198.64 | 0.17 |
| | LocalInfo | 3612.19 | 3114.46 | 0.16 | 3770.78 | 3198.64 | 0.18 |
| | Freesurfer | 5018.76 | 3114.46 | 0.61 | 4730.86 | 3198.64 | 0.48 |
| 3 | ABSS | 2086.07 | 2724.09 | 0.23 | 2526.46 | 1950.66 | 0.30 |
| | LocalInfo | 3469.46 | 2724.09 | 0.27 | 2581.35 | 1950.66 | 0.32 |
| | Freesurfer | 4458.81 | 2724.09 | 0.64 | 3376.74 | 1950.66 | 0.73 |
| 4 | ABSS | 3447.50 | 2819.24 | 0.22 | 2765.56 | 3229.13 | 0.14 |
| | LocalInfo | 4056.24 | 2819.24 | 0.44 | 4114.80 | 3229.13 | 0.27 |
| | Freesurfer | 4360.00 | 2819.24 | 0.55 | 4523.47 | 3229.13 | 0.40 |

**Table 2:** Sample of 426 patient samples comparing the left and right volumes of the hippocampus by automatic and manual segmentation.

| Measurement | Source | SS | df | MS | F | P-value |
|---|---|---|---|---|---|---|
| Dice | Columns | 2.34 | 2 | 1.17 | 118.37 | 7.18E-48 |
| | Error | 12.56 | 1275 | 0.01 | - | - |
| | Total | 14.90 | 1277 | - | - | - |
| Haudroff | Columns | 2168.30 | 2 | 1084.14 | 44.42 | 2.60E-19 |
| | Error | 31046.80 | 1275 | 24.41 | - | - |
| | Total | 33215.10 | 1277 | - | - | - |
| Percision | Columns | 7.59 | 2 | 3.79 | 282.11 | 3.95E-102 |
| | Error | 17.11 | 1275 | 0.01 | - | - |
| | Total | 24.70 | 1277 | - | - | - |
| RAVD | Columns | 73.45 | 2 | 36.72 | 194.57 | 1.88E-74 |
| | Error | 240.08 | 1275 | 0.19 | - | - |
| | Total | 313.52 | 1277 | - | - | - |
| RMS | Columns | 248.40 | 2 | 124.20 | 29.57 | 2.81E-13 |
| | Error | 5343.14 | 1275 | 4.20 | - | - |
| | Total | 5591.55 | 1277 | - | - | - |
| Sensitivity | Columns | 1.06 | 2 | 0.53 | 56.34 | 3.59E-24 |
| | Error | 11.98 | 1275 | 0.01 | - | - |
| | Total | 13.04 | 1277 | - | - | - |
| Similarity | Columns | 3.06 | 3.1095 | 1.53 | 147.06 | 3.49E-59 |
| | Error | 13.25 | 1275 | 0.01 | - | - |
| | Total | 16.31 | 1277 | - | - | - |
| ASSD | Columns | 110.22 | 169.85 | 55.11 | 28.79 | 5.93E-13 |
| | Error | 2435.24 | 1275 | 1.91 | - | - |
| | Total | 2545.46 | 1277 | - | - | - |

**Table 3:** Standard ANOVA table using data from 426 patients and three automatic segmentation methods. The source of the variability is shown in the source column. The sum of squares (SS) from each of the sources is depicted in the third column and the degrees of freedom (df) in the fourth. The fifth is the mean of squares (MS), which is the ratio of the sum of the squares/degrees of freedom. The sixth column is the F-statistic (F), the ratio of the mean squares, and the last column is the p-value from the cumulative distribution of F.

The ABSS method outperforms the other two with all eight similarity measures showing superiority. Moreover, RAVD is greater than zero for all three methods suggesting the segmented hippocampi are larger than the actual size. In other words, the estimated volumes are biased to larger values. However, the segmented volume by ABSS is the closest to the actual volume while the one by FreeSurfer is the furthest. The segmentation performance of LocalInfo stays between the two other methods but closer to FreeSurfer.

Performance measures are altered by field inhomogeneity induced by the magnet in the 1.5T and 3T MRI. For example, the Dice coefficient and precision values of FreeSurfer improve slightly when using the data of 3T MRI. On the other hand, these performance measures decrease slightly for ABSS when 3T MRI is used. Also, the Hausdorff distance of all three methods increase when using 3T MRI data.

## 4. CONCLUSION

Manual segmentation of the hippocampus remains the gold standard for its accuracy in undertaking any metrics related to aging or disease. It is, however, a very time- consuming process. Automatic segmentations methods have been developed to expedite and standardize the process for greater reproducibility, but sample sizes to-date have been small. In this work, three automatic segmentation methods were evaluated using a dataset of 426 subjects including T1-weighted MRI. Eight performance measurements were applied to assess agreement between the automatic and manual hippocampal segmentations. We found ABSS to outperform other methods based on ASSD, Hausdorff distance, root mean square, similarity, dice coefficient, sensitivity, precision, Hausdorff, and volume agreement.

## REFERENCES

[1] A. Obenaus, "Neuroimaging Biomarkers for Epilepsy: Advances and Relevance to Glial Cells," *Neurochemistry international,* vol. 63, no. 7, pp. 712-718, 05/09, 2013.

[2] M. Chupin, E. Gérardin, R. Cuingnet, C. Boutet, L. Lemieux, S. Lehéricy, H. Benali, L. Garnero, O. Colliot, and I. and the Alzheimer's Disease Neuroimaging, "Fully Automatic Hippocampus Segmentation and Classification in Alzheimer's Disease and Mild Cognitive Impairment Applied on Data from ADNI," *Hippocampus,* vol. 19, no. 6, pp. 579-587, 2009.

[3] M. P. Hosseini, M. R. Nazem-Zadeh, D. Pompili, K. Jafari-Khouzani, K. Elisevich, and H. Soltanian-Zadeh, "Comparative performance evaluation of automated segmentation methods of hippocampus from magnetic resonance images of temporal lobe epilepsy patients," *Med Phys,* vol. 43, no. 1, pp. 538, Jan, 2016.

[4] B. Fischl, D. H. Salat, E. Busa, M. Albert, M. Dieterich, C. Haselgrove, A. van der Kouwe, R. Killiany, D. Kennedy, S. Klaveness, A. Montillo, N. Makris, B. Rosen, and A. M. Dale, "Whole brain segmentation: automated labeling of neuroanatomical structures in the human brain," *Neuron,* vol. 33, no. 3, pp. 341-55, Jan 31, 2002.

[5] D. Shen, and C. Davatzikos, "HAMMER: hierarchical attribute matching mechanism for elastic registration," *IEEE Trans Med Imaging,* vol. 21, no. 11, pp. 1421-39, Nov, 2002.

[6] M. J. *Moghaddam*, and H. Soltanian-Zadeh, "Automatic Segmentation of Brain Structures Using Geometric Moment Invariants and Artificial Neural Networks," in Proceedings of the 21st International Conference on Information Processing in Medical Imaging, Williamsburg, Virginia, 2009, pp. 326-337.

[7] A. Akhondi-Asl, K. Jafari-Khouzani, K. Elisevich, and H. Soltanian-Zadeh, "Hippocampal volumetry for *lateralization* of temporal lobe epilepsy: Automated versus manual methods," *NeuroImage,* vol. 54, Supplement 1, pp. S218-S226, 2011.

[8] Despotovi, #x107; I. , B. Goossens, and W. Philips, "MRI Segmentation of the Human Brain: Challenges, Methods, *and* Applications," *Computational and Mathematical Methods in Medicine,* vol. 2015, pp. 23, 2015.

[9] Dill, V., Franco, A. R., & Pinho, M. S. (2015). Automated methods for hippocampus segmentation: the evolution and a review of the state of the art. *Neuroinformatics*, 13(2), 133-150.

[10] N. Nogovitsyn, Roberto Souza, M. Muller, and Amerlia Srajer, "Testing a deep convolutional neural network for automated hippocampus segmentation in a longitudinal sample of healthy participants," *NeuroImage*, vol. 197, pp. 589–597, Aug. 2019.

[11] Frisoni, G. B., Jack Jr, C. R., Bocchetta, M., Bauer, C., Frederiksen, K. S., Liu, Y., ... & Grothe, M. J. (2015). The EADC-ADNI Harmonized Protocol for manual hippocampal segmentation on magnetic resonance: evidence of validity. *Alzheimer's & Dementia*, 11(2), 111-125.

[12] S. Pang, Z. Lu, J. Jiang, L. Zhao, X. Li, M. Huang, W. Yang, and Q. Feng, "Hippocampus Segmentation based on Iterative Local Linear Mapping with Representative and Local Structure-preserved Feature Embedding," *IEEE Transactions on Medical Imaging*, pp. 1–1, 2019.

[13] Bartel, F., Vrenken, H., van Herk, M., de Ruiter, M., Belderbos, J., Hulshof, J., & de Munck, J. C. (2019). FAst Segmentation Through SURface Fairing (FASTSURF): A novel semi-automatic hippocampus segmentation method. *PloS one*, 14(1), e0210641.

[14] S. Hurtz, N. Chow, A. E. Watson, J. H. Somme, N. Goukasian, K. S. Hwang, J. Morra, D. Elashoff, S. Gao, R. C. Petersen, P. S. Aisen, P. M. Thompson, and L. G. Apostolova, "Automated and manual hippocampal segmentation techniques: Comparison of results, reproducibility and clinical applicability," *NeuroImage: Clinical*, vol. 21, p. 101574, 2019.

[15] Chupin, M., Hammers, A., Liu, R. S., Colliot, O., Burdett, J., Bardinet, E., ... & Lemieux, L. (2009). Automatic segmentation of the hippocampus and the amygdala driven by hybrid constraints: method and validation. *Neuroimage,* 46(3), 749-761.

[16] Y. L. Fung, K. E. Ng, S. J. Vogrin, C. Meade, M. Ngo, S. J. Collins, and S. C. Bowden, "Comparative Utility of Manual versus Automated Segmentation of Hippocampus and Entorhinal Cortex Volumes in a Memory Clinic Sample," *Journal of Alzheimers Disease*, vol. 68, no. 1, pp. 159–171, 2019.

[17] Khlif, M. S., Egorova, N., Werden, E., Redolfi, A., Boccardi, M., DeCarli, C. S., ... & Brodtmann, A. (2019). A comparison of automated segmentation and manual tracing in estimating hippocampal volume in ischemic stroke and healthy control participants. *NeuroImage: Clinical*, *21*, 101581.

[18] Cover, K. S., van Schijndel, R. A., Versteeg, A., Leung, K. K., Mulder, E. R., Jong, R. A., ... & Manset, D. (2016). Reproducibility of hippocampal atrophy rates measured with manual, FreeSurfer, AdaBoost, FSL/FIRST and the MAPS-HBSI methods in Alzheimer's disease. *Psychiatry Research: Neuroimaging*, *252*, 26-35.

[19] Maglietta, R., Amoroso, N., Boccardi, M., Bruno, S., Chincarini, A., Frisoni, G. B., ... & Bellotti, R. (2016). Automated hippocampal segmentation in 3D MRI using random undersampling with boosting algorithm. *Pattern Analysis and Applications*, *19*(2), 579-591.

[20] Morey, R. A., Petty, C. M., Xu, Y., Hayes, J. P., Wagner II, H. R., Lewis, D. V., ... & McCarthy, G. (2009). A comparison of automated segmentation and manual tracing for quantifying hippocampal and amygdala volumes. *Neuroimage*, *45*(3), 855-866.

[21] Fraser, M. A., Shaw, M. E., Anstey, K. J., & Cherbuin, N. (2018). Longitudinal assessment of hippocampal atrophy in midlife and early old age: contrasting manual tracing and semi-automated segmentation (FreeSurfer). *Brain topography*, *31*(6), 949-962.

[22] Zandifar, A., Fonov, V., Coupé, P., Pruessner, J., Collins, D. L., & Alzheimer's Disease Neuroimaging Initiative. (2017). A comparison of accurate automatic hippocampal segmentation methods. *NeuroImage*, *155*, 383-393.

[23] Schoemaker, D., Buss, C., Head, K., Sandman, C. A., Davis, E. P., Chakravarty, M. M., ... & Pruessner, J. C. (2016). Hippocampus and amygdala volumes from magnetic resonance images in children: Assessing accuracy of FreeSurfer and FSL against manual segmentation. *NeuroImage*, *129*, 1-14.

[14] M. L. Schlichting, M. L. Mack, K. F. Guarino, and A. R. Preston, "Performance of semi-automated hippocampal subfield segmentation methods across ages in a pediatric sample," *NeuroImage*, vol. 191, May 2019.

[25] Wenger, E., Mårtensson, J., Noack, H., Bodammer, N. C., Kühn, S., Schaefer, S., ... & Lövdén, M. (2014). Comparing manual and automatic segmentation of hippocampal volumes: reliability and validity issues in younger and older brains. *Human brain mapping*, *35*(8), 4236-4248.

[26] Colliot, O., Chételat, G., Chupin, M., Desgranges, B., Magnin, B., Benali, H., ... & Lehéricy, S. (2008). Discrimination between Alzheimer disease, mild cognitive impairment, and normal aging by using automated segmentation of the hippocampus. *Radiology*, *248*(1), 194-201.

[27] K. Jafari-Khouzani, K. V. Elisevich, S. Patel, and H. Soltanian-Zadeh, "Dataset of magnetic resonance images of nonepileptic subjects and temporal lobe epilepsy patients for validation of hippocampal segmentation techniques," *Neuroinformatics,* vol. 9, no. 4, pp. 335-46, Dec, 2011.

[28] "https://sites.google.com/site/jabaroutimoghaddam/."

[29] "http://freesurfer.net/."

[30] Ataloglou, D., Dimou, A., Zarpalas, D., & Daras, P. (2019). Fast and precise hippocampus segmentation through deep convolutional neural network ensembles and transfer learning. *Neuroinformatics*, *17*(4), 563-582.

[31] Hosseini, M. P., Nazem-Zadeh, M. R., Pompili, D., & Soltanian-Zadeh, H. (2014, August). Statistical validation of automatic methods for hippocampus segmentation in mr images of epileptic patients. In *2014 36th Annual International Conference of the IEEE Engineering in Medicine and Biology Society* (pp. 4707-4710). IEEE.

[32] Hosseini, M. P., Nazem-Zadeh, M. R., Pompili, D., Jafari-Khouzani, K., Elisevich, K., & Soltanian-Zadeh, H. (2016). Automatic and manual segmentation of hippocampus in epileptic patients mri. *arXiv preprint arXiv:1610.07557*.

[33] Hosseini, M. P., Lau, A., Elisevich, K., & Soltanian-Zadeh, H. (2019). Multimodal Analysis in Biomedicine. *Big Data in Multimodal Medical Imaging*, 193.

[34] Hosseini, M. P., Lu, S., Kamaraj, K., Slowikowski, A., & Venkatesh, H. C. (2020). Deep Learning Architectures. In *Deep Learning: Concepts and Architectures* (pp. 1-24). Springer, Cham.

[35] Hosseini, M. P., Tran, T. X., Pompili, D., Elisevich, K., & Soltanian-Zadeh, H. (2020). Multimodal data analysis of epileptic EEG and rs-fMRI via deep learning and edge computing. *Artificial Intelligence in Medicine*, *104*, 101813.